\documentclass[%
 reprint,
nofootinbib,
 amsmath,amssymb,
 aps,
]{revtex4-2}

\usepackage{graphicx}
\usepackage{xcolor}
\usepackage{dcolumn}
\usepackage{bm}
 \usepackage[utf8]{inputenc}
\usepackage{mathtools}
\DeclarePairedDelimiter{\ceil}{\lceil}{\rceil}

\newcommand{\bfE}{{\bf{E}}}
\newcommand{\bfB}{{\bf{B}}}

\newcommand{\bfx}{{\bf{x}}}

\newcommand{\bfJ}{{\bf{J}}}

\newcommand{\bfOmega}{{\boldsymbol{\Omega}}}

\newcommand{\bfmu}{{\boldsymbol{\mu}}}

\newcommand{\gagg}{g_{a\gamma \gamma}}

\begin{document}

\preprint{APS/123-QED}

\title{Axion Production in Pulsar Magnetosphere Gaps}

\author{Anirudh Prabhu}
\email{aniprabhu@stanford.edu}
 \affiliation{Stanford Institute for Theoretical Physics, Stanford University, Stanford, California 94305, USA}

\date{\today}

\begin{abstract}

Pulsar magnetospheres admit non-stationary vacuum gaps that are characterized by non-vanishing ${\bf{E}} \cdot {\bf{B}}$. The vacuum gaps play an important role in plasma production and electromagnetic wave emission. We show that these gaps generate axions whose energy is set by the gap oscillation frequency. The density of axions produced in a gap can be several orders of magnitude greater than the ambient dark matter density. In the strong pulsar magnetic field, a fraction of these axions may convert to photons, giving rise to broadband radio signals. We show that dedicated observations of nearby pulsars with radio telescopes (FAST) and interferometers (SKA) can probe axion-photon couplings that are a few orders of magnitude lower than current astrophysical bounds.
\end{abstract}

\maketitle


\section{\label{sec:intro}Introduction}

Despite the many successes of the Standard Model (SM), a number of observations point to the need for new physics. Limits on the neutron electric dipole moment (EDM) \cite{EDM} suggest a finely tuned cancellation of the topological $\theta$ parameter in QCD and the overall phase of the quark mass matrix, a fine-tuning constraint known as the Strong CP problem. In addition, a number of cosmological observations point to an abundance of dark matter that does not appear to consist of SM particles. One of the best-motivated solutions to both of these problems is the QCD axion, which is the pseudo Nambu-Goldstone boson of the spontaneously broken Peccei-Quinn symmetry \cite{PQ1, PQ2, WeinbergAxion, WilczekAxion}. Non-perturbative effects generate a potential for the QCD axion that dynamically relaxes the $\theta$ parameter to its CP-conserving value, explaining the smallness of the neutron (EDM). In addition, QCD axions may be produced in the correct abundance to account for dark matter \cite{Abbott1982, Fischler1982, PRESKILL1983127}. Axions also appear more broadly in string theory in the compactification of antisymmetric tensor fields on Calabi-Yau manifolds \cite{Svrcek_2006,Axiverse2010}. The experimental discovery of axions would be an important step towards understanding physics beyond the SM. 

A number of direct detection experiments have been proposed to detect axions by exploiting their coupling to photons, $\mathcal{L} \supset -\gagg a(x) {\bfE \cdot \bfB}$. Some of these experiments make use of high-$Q$ microwave cavities \cite{Sikivie1983, Bradley2003, Bradley2004, Asztalos2010, Du2018, HAYSTAC, mcallister2017organ}, gapped toroidal magnets \cite{ABRACADABRA}, and dielectric  haloscopes \cite{MADMAX, MashaCurlyRobert2018}. In the aforementioned experiments, axions are {assumed to be} produced in the early universe and make up a significant portion of dark matter. {Other experiments do not rely on axions being dark matter}. {For example, axions may be} produced by the Primakoff process in the Sun \cite{CAST2009, CAST2015} or by an oscillating background $\bfE \cdot \bfB$ {field} in the laboratory \cite{LSW1987, LSW1992, LSW2007, LSW20072, LSW2009, CROWS, OSQAR2015}. When produced in the laboratory by photons with incident $\bfE$ parallel to an externally applied $\bfB$, axions may penetrate optical barriers and convert back to photons giving the appearance of light shining through walls (LSW). LSW experiments are not currently competitive with astrophysical bounds on axions, due in large part to their reliance on two low-probability events, photon-to-axion conversion and axion-to-photon conversion. Signal power is additionally suppressed by the limitations on the strength of magnetic fields that can be produced in the laboratory. 

Beyond the laboratory, astrophysical settings are often ideal testbeds for theories of physics beyond the SM. Neutron stars, in particular, provide excellent laboratories for axion detection due to their enormous magnetic fields and dense plasma magnetospheres. Infalling axion dark matter may resonantly convert to photons in neutron star magnetospheres to produce thin lines that may be detectable with current and planned radio missions \cite{Anson2018, Safdi2019, leroy2019, battye2020, foster2020, fapeng2020, Rapidis2020}. Another interesting possibility is that neutron stars and white dwarfs may be factories for axions. In particular, axions may be thermally produced in the cores of extreme astrophysical objects and converted to photons in their magnetospheres \cite{Kuver20181, Kuver20182, Dessert2019, DessertWD2019, Kuver2020, Kuver2021}. In this paper we discuss a new mechanism for axion production by pulsars. Pulsars, young, rapidly-rotating neutron stars, are surrounded by a dense plasma of electrons, positrons, and ions which are thought to be responsible for gamma ray, X-ray, and radio emission. In response to plasma outflow along open magnetic field lines, the magnetosphere forms non-stationary vacuum gaps in which $\bfE \cdot \bfB \ne 0$ \cite{Sturrock1971a, RudermanSutherland1975}. The oscillating $\bfE \cdot \bfB$ {in a gap} acts as a source for axions, {similar} to axion generation in LSW experiments. The energy density of axions produced by {a} gap can be several orders of magnitude higher than the corresponding dark matter energy density, and hence {can} give rise to very bright radio signals.

The paper is organized as follows. In Section \ref{sec:pulsars}, we review some basics of pulsar magnetospheres, including the formation of non-stationary vacuum gaps for plasma regeneration. In Section \ref{sec:production}, we present a concrete model for a gap, based on which we compute the spectrum of the axions produced in {that} gap. In Section \ref{sec:conversion} we discuss the conversion of axions to radio photons in the strong magnetic fields surrounding the pulsar. In Section \ref{sec:observation} we discuss an observation scheme using existing and planned radio telescopes, with a discussion of instrumental and astrophysical backgrounds. We conclude this section with an estimate of the reach of our scheme. We end with some concluding remarks in Section \ref{sec:conclusions}.

\section{\label{sec:pulsars} Theory of Pulsar Magnetospheres}

\color{black}
Pulsars are highly magnetized compact stars that emit {electromagnetic radiation} over a wide range of frequencies. Many pulsars have been observed due to their coherent, pulsed emission of radio waves. While the mechanism for coherent radio emission remains a mystery, it is widely believed that the dense electron-positron plasma that fills the magnetosphere plays an important role. Early explanations include coherent radio emission by dense, charged bunches \cite{RudermanSutherland1975} and maser-like two-stream plasma instabilities \cite{BlandfordTwoStream1999}. The explanation based on emission by charged bunches has fallen out of favor on theoretical grounds as a result of the difficulty in explaining the origin and longevity of these bunches \cite{Melrose1995} {as well as} on observational grounds \cite{Lesch1998}. A more recent proposal claims that non-uniform plasma generation in the magnetosphere, to be discussed subsequently, {gives rise to} electromagnetic radiation {whose} power and spectrum {are in agreement with} observations \cite{Spitkovsky2020}. We show that the processes responsible for plasma generation are also very efficient axion factories.

The presence of plasma is essential not only to the emission of coherent radio waves, but also to the stability of the magnetosphere \cite{GoldreichJulian1969}. Generation of electron-positron pairs relies on the formation of charge-starved ``gaps'' in the magnetosphere, which arise as a result of plasma outflow, along open field lines, through the light cylinder. The gap possesses a large component $\bfE_\parallel$, parallel to the magnetic field, that accelerates charged particles to high energy. Once sufficient charge generation has taken place, $\bfE_\parallel$ is screened and the gap {is} fully discharged. {As a result, large amplitude fluctuations of $\bfE \cdot \bfB$ occur in the gap}. The formation of gaps and their subsequent discharge has been demonstrated in numerical simulations \cite{TimokhinArons2013}. 
\color{black}


{A pulsar is} described by its angular velocity $\bfOmega$ and magnetic moment $\bfmu$. We limit our discussion to aligned rotators ($\bfmu \parallel \bfOmega$). While aligned rotators cannot describe the observed pulsation characteristic of pulsars, they capture the important details of pulsar dynamics. The surface of a NS is an excellent electric conductor. Therefore the {surface} electric field measured in the co-rotating frame of a NS is zero. In the laboratory frame, electrons redistribute themselves to counteract the Lorentz force due to the internal magnetic field yielding the following (equivalent) surface condition

\begin{align}
    \bfE_{\text{int}} + {\bf v} \times \bfB_{\text{int}} = 0 \label{eqn:intcond}
\end{align}

\noindent where ${\bf v} = \bfOmega\times {\bf r}$. For a vacuum magnetosphere (\ref{eqn:intcond}) may be used to find the external electrostatic potential,

\begin{align}
    \Phi(r, \theta) = -{B_0 \Omega R^5 \over 3 r^3} P_2\left(\cos\theta \right) \label{eqn:extpotential}
\end{align}

\noindent where $B_0$ is the {magnitude of the} surface magnetic field, $R$ is the NS radius, and $P_n(x)$ are Legendre polynomials. An important consequence of (\ref{eqn:extpotential}) is that there exists an external electric field component, $\bfE_\parallel$, parallel to the dipolar magnetic field. It was shown by Goldreich and Julian \cite{GoldreichJulian1969} that the vacuum magnetosphere solution is unstable. The electric force on charges due to $\bfE_\parallel$ dominates gravity and strips charges from the NS surface, accelerating them along magnetic field lines and populating the magnetosphere. Eventually, the plasma surrounding the NS screens $\bfE_\parallel$ and co-rotates with the pulsar. The co-rotating plasma charge density is given by \cite{GoldreichJulian1969} 

\begin{align}
    \rho_c = {2 \bfOmega \cdot \bfB \over 1 - \Omega^2 r^2 \sin^2\theta}. \label{eqn:GJdensity}
\end{align}

Co-rotation cannot be sustained at a radial distance (defined in cylindrical coordinates) $r > R_{LC} = c/\Omega$ where $R_{LC}$ is the radius of the light cylinder. The light cylinder defines two categories of magnetic field lines: \emph{closed} and \emph{open}, with the former being entirely contained within the light cylinder and the latter exiting it. Particles moving along open field lines may escape freely to infinity. {As a result, the region containing the open field lines lacks the charges necessary to screen $\bfE_\parallel$, leading to the formation of gaps}. While the existence of pulsar gaps has been well-established, there is some uncertainty about the mechanism for generating $\bfE_\parallel$ and the location of gaps. Various gap models have been proposed including the polar-cap \cite{Sturrock1971a, RudermanSutherland1975}, slot gap \cite{Arons1983, Muslimov2004}, and outer gap models \cite{Cheng1986}. In this article we focus on the observable signatures of polar-cap gap models.

\subsection{Polar Cap Production}

Along open field lines co-rotation cannot be sustained. {As a result, the electric potential deviates from that given by (\ref{eqn:extpotential}) and the plasma that lies on open field lines decelerates relative to the NS.} Additionally, plasma can leave the light cylinder through open field lines, leading to a violation of the longitudinal electric field screening condition in a gap right above the polar cap (the region on the NS surface out of which open field lines flow). The longitudinal voltage drop across the gap can exceed $10^{12}$ V \cite{Gurevich1993}. The strong longitudinal electric field in this gap is unstable to $e^+e^-$ plasma generation. {Repeated} formation and discharge of the vacuum gap is responsible for pulsar activity and, as {we show below}, the production of axions. In the {ensuing} discussion we follow the concrete model developed by Ruderman and Sutherland \cite{RudermanSutherland1975} (hereafter referred to as the RS model)\footnote{The RS model is restricted to anti-aligned pulsars where $\bfOmega \cdot \bfB < 0$ and the plasma above the polar cap is positively charged. The mechanism has been extended to aligned pulsars in, for example, \cite{Gil2003}.}. In the RS model, positive charges flow out of the light cylinder along open field lines and cannot be easily replenished by either the charge-separated magnetosphere or the NS surface, which strongly binds positively charged ions. {Consequently, a gap of thickness $h\ll R$ forms} above the polar cap. The potential drop across the polar gap is $\Delta V = \Omega^* B h^2$ where $\Omega^*$ is the rotational velocity of plasma in the open magnetosphere, which is the same as the rotational velocity of the NS up to $\mathcal{O}(h^2/R^2)$. 

\subsection{Dynamics of the Gap}

The gap width $h$ grows at roughly the speed of light with the potential drop across the gap growing as $h^2$. Gamma rays with energy $\omega \gg 2 m_e$ may convert to $e^+e^-$ pairs in the strong gap fields; the $e^+e^-$ pairs travel along curved field lines and give off curvature radiation which can further pair-produce in the electric and magnetic fields. Pair-production proceeds until the gap width exceeds the mean free path of energetic gamma rays with $\omega > 2 m_e$, at which point the gap breaks down into a {series of} sparks. This condition occurs when \cite{RudermanSutherland1975}

\begin{align}
    h \approx 360 \text{ cm} \left( \rho \over 10 \text{ km} \right)^{2/7} \left(\Omega \over \text{Hz} \right)^{-3/7} \left(B_s \over 10^{14} \text{ G}\right)^{-4/7}, \label{eqn:h}
\end{align}

\noindent where $\rho$ is the radius of curvature of field lines exiting the polar cap and $B_s$ is the magnitude of the surface magnetic field. The potential drop across the gap vanishes once $e^+e^-$ density coincides with (\ref{eqn:GJdensity}) and $\bfE_\parallel$ can be effectively screened. To capture these broad features, we assume $\bfE \cdot \bfB$ is periodic in time (with period $T$) with 

\begin{align}
    &\bfE \cdot \bfB(\bfx, t) =  \nonumber \\
    &2 \Omega B_s^2 h \left\{\begin{array}{ccc}
        t/h & 0 \le z \le t, & 0 \le t \le h  \\
        1  &  0 \le z \le h, & h < t \le T - h \\
        \left( T - t \over h \right) & z \le 0 \le T - t, & T-h < t \le T
    \end{array} \right., \label{eqn:gapdynamics}
\end{align}

\noindent and the radial distance from the rotation axis is less than the polar cap radius, $r_{\text{pc}}$. Here $z$ represents the direction normal to the NS surface at the polar cap. {Before discussing axion production in this model, we highlight a few effects that should be considered in a more sophisticated model.} Firstly, the mean free path of the aforementioned gamma rays depends exponentially on the magnetic field and gamma ray energy. Therefore, small fluctuations in the magnetic field can lead to inhomogeneous spark formation in the gap and a complex sub-burst structure. These fluctuations, however, do not change either the axion density or frequency spectrum by more than $\mathcal{O}(1)$ fractional amounts and will thus be neglected in the following discussion. {Secondly, it is unlikely that gap formation and discharge occurs periodically. More realistically, there will be variations in the time between successive discharge events. We discuss the relaxation of the periodicity assumption in the following section.}

The periodic time oscillation of the gap produces axions with energy given by the Fourier modes of the oscillation,

\begin{subequations}
\begin{equation}
    (\bfE \cdot \bfB) (\bfx, t) = \displaystyle\sum_{n} (\bfE \cdot \bfB)_n (\bfx) e^{i \omega_n t} \label{eqn:fouriera}
    \end{equation}
    \begin{equation}
    (\bfE \cdot \bfB)_n (\bfx) = {2\over T} \displaystyle\int_0^T (\bfE \cdot \bfB) (\bfx, t) e^{i \omega_n t}. \label{eqn:fourierb}
    \end{equation}
\end{subequations}

\noindent where $\omega_n = 2\pi n/T$. Equation (6) {represents the axion spectral content}. For some standard pulsar parameters these frequencies are

\begin{align}
    \frac{\omega_1}{2\pi} = 42 \text{ MHz} \left( \frac{B_s}{10^{14} \text{ G}} \right)^{4/7} \left( \frac{\Omega}{\text{ Hz}} \right)^{3/7} \left( \frac{c T}{2 h} \right)^{-1}.
\end{align}

\color{black} The time-scale of oscillation here is set by the crossing time of the gap at maximum thickness. Another time-scale which may contribute to axion production is the plasma frequency near the pulsar surface. \color{black}

\begin{figure}
    \centering
    \includegraphics[scale=0.375]{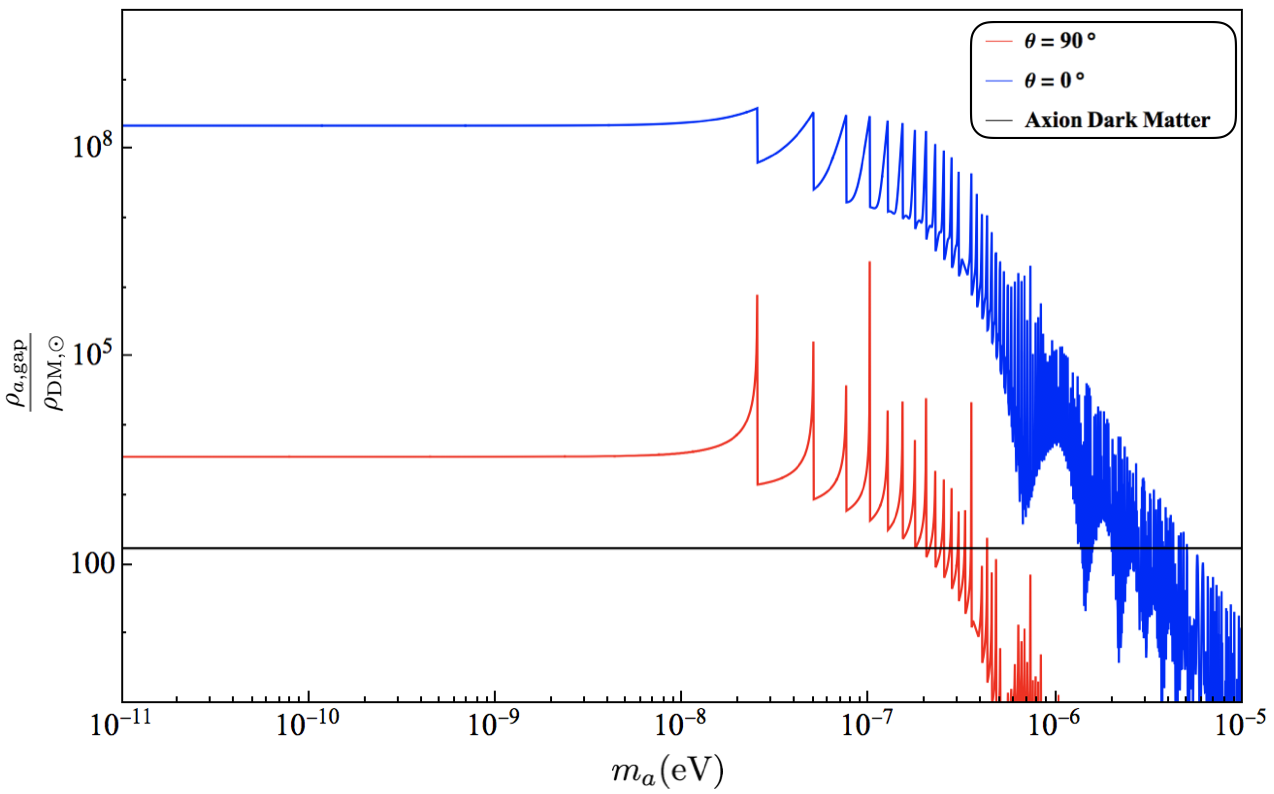}
    \caption{Comparison of axion density produced by gap oscillation to that of infalling axion dark matter at a distance of 100 km from the NS surface. The comparison is made for pulsar J1856--3754 at angles $\theta = 0^\circ$ (blue) and $\theta = 90^\circ$ (red) with respect to the rotation axis. The axion dark matter density infalling onto the NS (black) is larger than the local DM density as a result of Liouville's theorem. The axion-photon coupling is taken to be $\gagg = 10^{-11}$ GeV$^{-1}$. }
    \label{fig:densitycomp}
\end{figure}

\section{Axion Production in Gaps} \label{sec:production}

\begin{figure}
     \centering
     \includegraphics[scale=0.37]{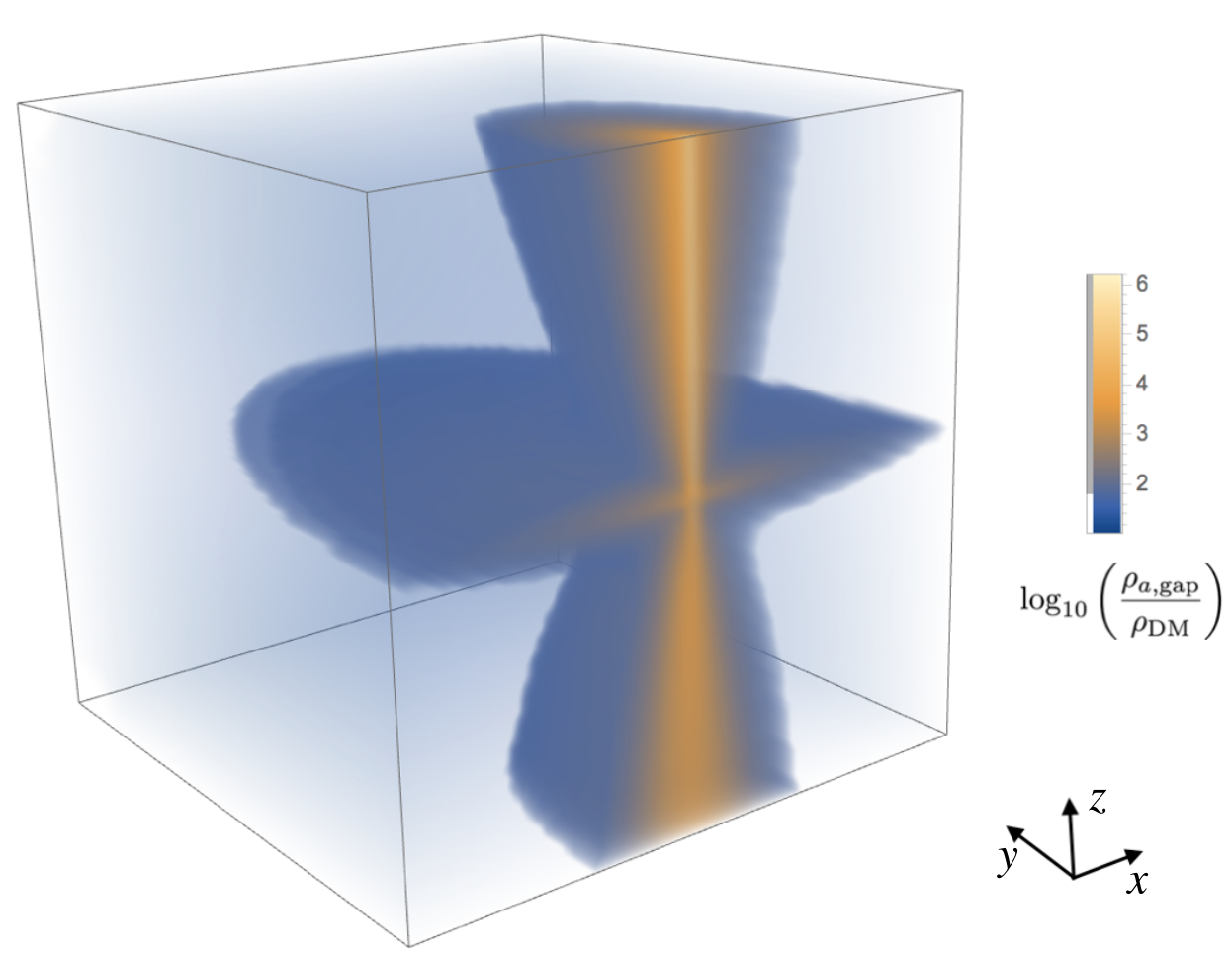}
     \caption{Comparison of the axion density produced by the gap to the dark matter density as a function of position in the magnetosphere. Axion emission is beamed in the direction normal to the polar cap ($z$-direction). The parameters are taken to be: $m_a = 10^{-7}$ eV, $\gagg = 10^{-11}$ GeV$^{-1}$, $B_0 = 10^{14}$ Gauss, $P= 1$ sec. }
     \label{fig:beaming}
 \end{figure}

Axions couple to EM fields through the effective Lagrangian,

\begin{align}
&\mathcal{L}_{\text{EM} + a} =  \\
&-{1\over 4} F_{\mu \nu} F^{\mu \nu} +{1 \over 2} (\partial_\mu a)^2 -{1 \over 2} m_a^2 a^2
-{\gagg \over 4} a F_{\mu \nu} \tilde{F}^{\mu \nu}, \nonumber 
\end{align}

\noindent where $a$ is the axion field, $m_a$ is the axion mass, $\gagg$ is the axion-photon coupling, and $\tilde{F}^{\mu \nu} = \varepsilon^{\mu\nu\alpha\beta} F_{\alpha \beta}/2$ is the dual field tensor. The dynamics of axion-photon coupling are governed by the following equations,

\begin{align}
    \nabla \cdot \bfE &= {\rho \over \epsilon_0} - \gagg \nabla a \cdot \bfB, \label{eqn:gauss} \\
    \nabla \times \bfB &= {\partial \bfE \over \partial t} + \mu_0 \bfJ + \gagg \left(\dot{a} \bfB - \nabla a \times \bfE \right) \label{eqn:ampere} \\
    \left( \Box + m_a^2 \right) a &= -\gagg \bfE \cdot \bfB, \label{eqn:axion}
\end{align}

\noindent where $\rho$ and $\bfJ$ are the charge and current densities. The source-free Maxwell equations remain unaltered by the presence of an axion. According to (\ref{eqn:axion}), a region where $\bfE \cdot \bfB \ne 0$, such as a pulsar gap, can source axions. The production of axions using a background $\bfE \cdot \bfB$ component is employed in Light Shining through Walls (LSW) and related experiments \cite{LSW1987, LSW1992, LSW2007, LSW20072, LSW2009}, Any Light Particle Search (ALPS) \cite{ALPS1,ALPS2,ALPS3}, and the CERN Resonant Weakly Interacting sub-eV Particle Search (CROWS) \cite{CROWS}. In experiments such as CROWS, axions are produced in a cavity in which $\bfE \cdot \bfB \ne 0$ where $\bfE$ comes from an excited cavity mode and $\bfB$ is an externally applied, static magnetic field. A single mode of frequency $\omega$ sources an axion field,

\begin{align}
    a_\omega(\bfx,t) &= -{\gagg \over 4\pi} e^{i \omega t} \displaystyle\int d^3 \bfx' {e^{i k_a |\bfx - \bfx'|} \over |\bfx - \bfx'|} \left(\bfE \cdot \bfB \right)_\omega (\bfx'), \label{eqn:LSW}
\end{align}

\noindent where $k_a = \sqrt{\omega^2 - m_a^2}$ is the axion momentum and $\left(\bfE \cdot \bfB \right)_\omega$ is the component of $\bfE \cdot \bfB$ with frequency $\omega$. For our purposes the integral is performed over the pulsar gap. In pulsar gaps $\bfE \cdot \bfB$ is time-dependent due to the outflow of plasma through the light cylinder and the discharge of the gap {induced by} pair production. The axion density computed using (\ref{eqn:axion}) and (\ref{eqn:gapdynamics}) is shown in Fig. \ref{fig:densitycomp} {at two different magnetosphere locations in pulsar J1856--3754}. The appearance of resonance peaks at $m_a = \omega_n$ is a result of the assumption of gap oscillation periodicity, which is likely not realized in realistic models. In calculations of the signal power that follow we do not consider the effects of the resonance peaks shown in Fig. \ref{fig:densitycomp}. The total energy released in gap discharge is predominantly in the form of ultra-relativistic $e^+e^-$ pairs. The fraction of the gap energy that is released in the form of axions is $f \sim (\gagg \Omega B_s h^2)^2 \ll 1$ and hence the production of axions does not affect standard pulsar processes. \color{black}
Axions produced in the vacuum gap are also highly beamed in the direction normal to the polar cap, as demonstrated in Fig. \ref{fig:beaming}. 
\color{black}
 
\section{Axion-Photon Conversion} \label{sec:conversion}

Once produced, axions can penetrate the dense plasma until they reach a distance at which photon conversion becomes kinematically allowed. In the presence of a strongly magnetized plasma, axion electrodynamics (\ref{eqn:gauss}--\ref{eqn:axion}) must be supplemented by strong QED terms, 

\begin{align}
    \mathcal{L}_{\text{QED}} \supset {\alpha^2 \over 90 m_e^4} \left[ (F_{\mu \nu} F^{\mu \nu})^2 + {7 \over 4} (F_{\mu \nu} \tilde{F}^{\mu \nu})^2 \right],
\end{align}

\noindent where $\alpha$ is the electromagnetic fine-structure constant and $m_e$ is the electron mass. Axion-photon conversion results from mixing of the axion with the (parity-odd) state of the photon, $A_\parallel$, whose polarization is parallel to the external magnetic field, as shown below,

\begin{align}
    \left[ \partial_z^2 + \omega^2 + 2 \omega^2 \begin{pmatrix} \Delta_\parallel(z) & \Delta_B(z) \\
    \Delta_B(z) & \Delta_a \end{pmatrix}\right]   \begin{pmatrix} A_\parallel \\
    a\end{pmatrix} &= 0 \nonumber \\
    \Delta_\parallel =  - {\omega_p^2(z) \sin^2\theta \over 2 (\omega^2 - \omega_p(z)^2 \cos^2{\theta})} + {14 \alpha^2 \over 45} {B(z)^2 \sin^2\theta \over m_e^4}& \label{eqn:mixing2}\\
    \Delta_B = {\gagg \omega B(z) \sin{\theta} \over 2(\omega^2 - \omega_p(z)^2 \cos^2{\theta})}, \ \Delta_a = -{m_a^2 \over 2 \omega^2},  \nonumber
\end{align}

\noindent where $B$ is the magnitude of the magnetic field, $\theta$ is the angle it makes with the $z$ direction, $\omega_p(z)$ is the spatially varying plasma frequency, $\omega$ is the frequency of the axion/photon fields. The refractive index in the direction parallel to the transverse magnetic field, $n_\parallel$ ($\Delta_\parallel$ + 1) receives a contribution from strong-field QED. This term may be neglected for non-relativistic axions as it does not modify the resonance condition even for the highest magnetic fields observed in magnetars \cite{Anson2018}, however it must be restored when considering possibly ultrarelativistic axions converting close to the pulsar surface. We have ignored the Faraday effect, which {can} mix the two transverse EM modes \cite{RaffeltStodolsky1988}. 

The conversion probability between axions and photons depends critically on the relationship between the axion mass and various pulsar parameters \cite{witte2021axionphoton}. When the maximal mixing condition,

\begin{align}
    {\omega_p^2 \sin^2{\theta} \over 1 - {\omega_p^2\over \omega^2} \cos^2{\theta}} - {28 \alpha^2 \over 45} {B^2 \sin^2\theta \over m_e^4} \omega^2 = m_a^2, \label{eqn:mixingcondition}
\end{align}

\noindent is satisfied, conversion between axions and photons is resonantly enhanced. Strong QED effects  prevent maximal mixing from being achieved for axions with frequency $\omega > (360\pi^2 \Omega^2 m_e^2/7e^2)^{1/4} \approx 1.6 \times 10^{-4}$ eV $(\Omega/\text{Hz})^{1/2}$ such as those produced thermally in the cores of NSs \cite{Kuver20181, Kuver20182, Dessert2019, Kuver2020,  Kuver2021} and white dwarfs \cite{DessertWD2019}. At lesser frequencies, strong QED effects do not contribute significantly to conversion.

The fundamental Fourier mode of gap oscillations is $\omega_1 \sim 10^{-7}$ eV, which means axions produced by such oscillations can convert to photons resonantly even if they are ultra-relativistic. We restrict our analysis to $m_a \lesssim 10^{-5}$ eV {since for heavier axions the axion dark matter density dominates that of the axions produced by gap oscillations}. For light axions ($m_a < \omega_p(R_{LC})$) resonant conversion does not occur in the magnetosphere. Intermediate mass axions ($\omega_p(R_{LC}) < m_a \ll \omega_1$) require some care, as resonant conversion occurs in the outer magnetosphere where magnetic fields are considerably weaker than near the NS surface. 

The length scale over which the magnetic field and plasma properties change is generally much larger than the de Broglie wavelength of the axions and photons generated by the gap. Therefore we may approximate $\omega^2 + \partial_z^2 \approx (\omega + k)(\omega - i \partial_z)$. As mentioned above, for light axions the axion and photon dispersion relations do not coincide within the light cylinder. However, since light axions are ultra-relativistic, both the photon and axion momenta are well-approximated by $k_{a,\gamma} \approx \omega$. This approximation breaks down when $m_a$ becomes comparable to the fundamental oscillation frequency of the gap. However at this mass the conversion occurs resonantly and hence $k_a = k_\gamma = \sqrt{\omega^2 - m_a^2}$. In both regimes, we may reduce (\ref{eqn:mixing2}) to a system of coupled first-order differential equations \cite{RaffeltStodolsky1988}. For ultra-relativistic axions the simplified equation is

\begin{align}
    \left[ i \partial_z + \omega + \omega \begin{pmatrix} \Delta_\parallel(z) & \Delta_B(z) \\
    \Delta_B(z) & \Delta_a \end{pmatrix} \right] \begin{pmatrix} 
    A_\parallel \\
    a
    \end{pmatrix} = 0. \label{eqn:mixlinear}
\end{align}

In the weak-mixing limit ($\Delta_B \ll \Delta_\parallel - \Delta_a$), (\ref{eqn:mixlinear}) may be solved to first-order using time-dependent perturbation theory, giving a conversion probability 

\begin{align}
    P_{a\to \gamma} = \omega^2 \left| \displaystyle\int_{z_\text{min}}^\infty dz' \ \Delta_B(z') e^{i \omega \left( \Delta_a z' - \int_{z_\text{min}}^{z'} dz'' \Delta_\parallel(z'') \right) } \right|^2.
\end{align}

\noindent where $z_{\text{min}}$ is the radius at which the axion energy is equal to the photon mass, making photon conversion kinematically possible. In the limit of maximal mixing, conversion occurs in a narrow region surrounding a critical radius, $z_c$ defined by $\omega_p(z_c) = m_a$. In this limit the dispersion relations of the axion and photon coincide, allowing us to make the following ansatz $A_\parallel(z,t) = \bar{A}_\parallel(z) e^{-i\omega t + i k z}, \ a(z,t) = \bar{a}(z) e^{-i\omega t + i k z}$. This ansatz, along with the WKB approximation lead to the following first-order equation \cite{Anson2018},

\begin{align}
    \left[ i \partial_z  + {\omega^2 \over k} \begin{pmatrix} \Delta_\parallel(z) - \Delta_a & \Delta_B(z) \\
    \Delta_B(z) & 0\end{pmatrix} \right] \begin{pmatrix} 
    A_\parallel \\
    a
    \end{pmatrix} = 0. \label{eqn:mixlinearNR}
\end{align}

\noindent where $k=\sqrt{\omega^2 - k_a^2}$. We emphasize that (\ref{eqn:mixlinearNR}) is only valid in a narrow region around the conversion radius, $z_c$. The amplitude of the photon field once it has left the conversion region {is modulated by} the varying plasma density {and asymptotically} $A_\parallel(\infty) \approx \sqrt{k/\omega} A_\parallel(r_c)$. The final conversion probability, which includes the modulation of outgoing photons is

\begin{align}
    &P_{a\to\gamma, \text{res}} = \nonumber \\
    &{\omega^3 \over k} \left| \displaystyle\int_{z_\text{min}}^\infty dz' \ \Delta_B(z') e^{i {\omega^2 \over k} \left( \Delta_a z' - \int_{z_\text{min}}^{z'} dz'' \Delta_\parallel(z'') \right) } \right|^2.
\end{align}

Although equations (\ref{eqn:mixlinear}) and (\ref{eqn:mixlinearNR}) were derived in different limits, they are {almost identical}. \color{black} The total radio flux seen on Earth is related to the conversion probability as,

\begin{align}
    S = {2 \over D^2} \rho_a(z_c) P_{a \to \gamma} v_c z_c^2,
\end{align}

\noindent where $\rho_a(z)$ is the axion density at distance $z$ from the NS surface, $z_c$ is the radius at which conversion takes place, $v_c$ is the axion velocity at the conversion radius, and $D$ is the distance to the pulsar.  
\color{black}

\section{Observation and Backgrounds} \label{sec:observation}

\color{black}

Periodicity of gap formation and discharge would lead to a series of narrow axion-induced radio emission lines at integer multiples of $\omega_1 = 2\pi/T$. In reality, the gap breakdown is both inhomogeneous within the gap and likely not perfectly periodic, as suggested by the complex sub-burst structure of pulsar emissions. While the assumption of periodicity, and hence the calculation above provides a good approximation for the axion flux produced by gap oscillations, the breakdown of this assumption necessarily changes the signal search strategy. We therefore adopt a broadband strategy for signal detection. Most radio telescopes employ a spectrometer that divides the passband into sub-bands of width $\delta \nu = \Delta\nu/N_\nu$ to gather spectral information about the source. Radio data will consist of a series power measurements in the $N_\nu$ frequency bins and $N_t$ time bins each of width $\delta t = T_\text{int}/N_t$. The power of the axion-induced signal in frequency bin $i$ and time bin $j$ is denoted by $S_\text{ax}(i,j)$. Under the null hypothesis the power measurement will receive contributions from astrophysical and receiver noise sources to be discussed below. To test the hypothesis of an axion-induced signal we construct a $\chi^2$ test statistic,

\begin{align}
    \chi^2 = 2 \displaystyle\sum_{i=1}^{N_\nu} \displaystyle\sum_{j=1}^{N_t} {S_\text{ax}(i,j)^2 \over \sigma_S(i)^2}
\end{align}

\noindent where $\sigma_S$ is the flux density error in each bin, which is related to the familiar system equivalent flux density (SEFD) by $\sigma_S (\nu) = \text{SEFD}$. {The} factor of $2$ comes from the number of linear polarization states being measured.

\color{black}

The relevant quantity to calculate the sensitivity of various radio telescopes to \color{black} broadband \color{black} signals from axion-photon conversion is the flux density $\Phi = 1/(D^2 \Delta \nu_{\text{sig}}) dP/d\Omega$ where $dP/d\Omega $ is the signal power emitted in the direction of Earth, $D$ is the distance of the pulsar from Earth, and $\Delta \nu_{\text{sig}}$ is the bandwidth of the signal. For a given radio telescope or array the minimum detectable flux density at unit signal-to-noise is 

\begin{align}
    S_\text{min} &= {\text{SEFD} \over \sqrt{n_\text{pol} \Delta \nu_\text{rec}  T_{\text{int}}}}
\end{align}

\noindent where SEFD is the system equivalent flux density, $n_\text{pol}$ is the number of polarization states present in the image, $\Delta \nu_\text{rec}$ is the telescope bandwidth, and $T_\text{int}$ is the total integration time. 

The SEFD is defined as the flux density of a source that would deliver the same amount of power to the receiver as conventional noise sources. It may be expressed as SEFD$= 2 T_\text{noise}/A_\text{eff}$ where $T_\text{noise}$ is the noise temperature, which receives contributions from instrumental as well as astrophysical sources, and $A_\text{eff}$ is the effective area of the telescope or array. For a single-dish telescope the effective area is related to the physical area by an $\mathcal{O}(1)$ aperature efficiency term and for an array with $N\ge 2$ elements, the effective area  is greater than that of a single telescope by a factor of $\sqrt{N(N-1)}$. In the following subsection we estimate the instrumental and astrophysical backgrounds that contribute to SEFD.

\subsection{Backgrounds}

\color{black}
Conventional noise sources in radio astronomy include radiometer losses, galactic and extragalactic radio sources, the CMB, atmospheric emission, and spillover radiation from terrestrial sources.
\color{black} The total noise temperature at frequency $\nu$ is then approximately $T_\text{noise}(\nu) = T_R(\nu) + T_\text{astro}(\nu)$ where $T_R$ is the receiver noise, $T_\text{astro}$ is the noise coming from astrophysical sources unrelated to the NS. The receiver noise depends on the particular radio telescope(s) being used and is discussed below. Astrophysical noise is dominated by anisotropic galactic synchrotron emission as demonstrated by Haslam et al. \cite{Haslam1, Haslam2, Haslam3}. The radio sky also receives contributions from an isotropic extragalactic component. While these sources differ in their spatial and spectral properties, the distinction is not important for our purposes. A recent measurement of the absolute sky temperature by the ARCADE 2 collaboration \cite{ARCADE2} shows a spectrum 

\begin{align}
    T_\text{astro} = \left(24.1 \pm 2.1 \right) \text{ K} \left( \nu \over 310 \text{ MHz}\right)^{-2.599 \pm 0.036}.
\end{align}

\color{black}

\subsubsection{Pulsar and Nebula Radio Emission}

Ideal targets for observing the effects of axion conversion are nearby pulsars with high magnetic fields and minimal pulsed radio emission. The absence of coherent radio emission is often associated with inefficient plasma generation in the polar cap, and hence inefficient axion production therein. However, there are simple explanations, consistent with polar cap pair and axion production, for the lack of observed radio emission from certain pulsars. For example, radio-quiet pulsars may be beamed away from Earth, but are still detectable through quasi-isotropic thermal surface emission of soft X-rays \cite{Brazier1999}. Additionally, while axion emission is strongly beamed in the direction normal to the polar cap, coherent radio emission can be generated at a considerable distance and angle from the polar axis \cite{Spitkovsky2020}. Other classes of targets include soft gamma repeaters (SGRs) and anomalous X-ray pulsars (AXPs), which are highly magnetic pulsars that are widely believed to be magnetars. While few SGRs and AXPs have been observed to emit pulsed radio waves \cite{Camilo2006}, many of them are characterized by their X-ray and gamma-ray emissions, with no observed radio counterpart. For the strong magnetic fields in SGRs and AXPs, the lack of radio emission may not be incompatible with efficient charge generation near the polar cap. Once primary pairs are produced in the gaps of SGRs and AXPs, the third-order QED process $\gamma \to \gamma \gamma$ may quench further pair production from curvature radiation and hence coherent radio emission \cite{Baring1998}. {One potential pitfall of SGRs and AXPs is that they possess magnetic fields large enough to produce $e^+e^-$ pairs without the need for gaps.} While the exact plasma generation mechanism in magnetars remains an open problem, it is possible that they possess vacuum gaps \cite{Thompson2008}. {In the table below we include the relevant properties of three candidate stars with ultra-strong magnetic fields and no observed radio emission, perhaps due to one of the reasons mentioned in the preceding discussion.} SGR properties are taken from \cite{Kaspi2014}. For the selected pulsars we assume radio emission from these sources is dominated by standard astrophysical noise and receiver noise of the antennas. 


\begin{center}
\begin{tabular}{|c|c|c|c| } 
 \hline
 Name, Ref & $B_s \text{ (Gauss)}$ & $P_0 \text{ (sec)}$ & Dist. (kpc) \\ 
 \hline 
 RX J1856.5--3754 \cite{van_Kerkwijk_2008} & $1.5 \times 10^{13}$ & 7.1 & 0.16 \\
 \hline
 SGR 1900+14 \cite{Mereghetti2006} & $7.0 \times 10^{14}$ & 5.2 & 12.5  \\ 
 \hline
 SGR 1806--20 \cite{SGR180620} & $2.1 \times 10^{15}$ & 7.6 & 13 \\
 \hline
\end{tabular} \label{tab:pulsars}
\end{center}

\begin{figure}[h!]
    \centering
    \includegraphics[scale=0.41]{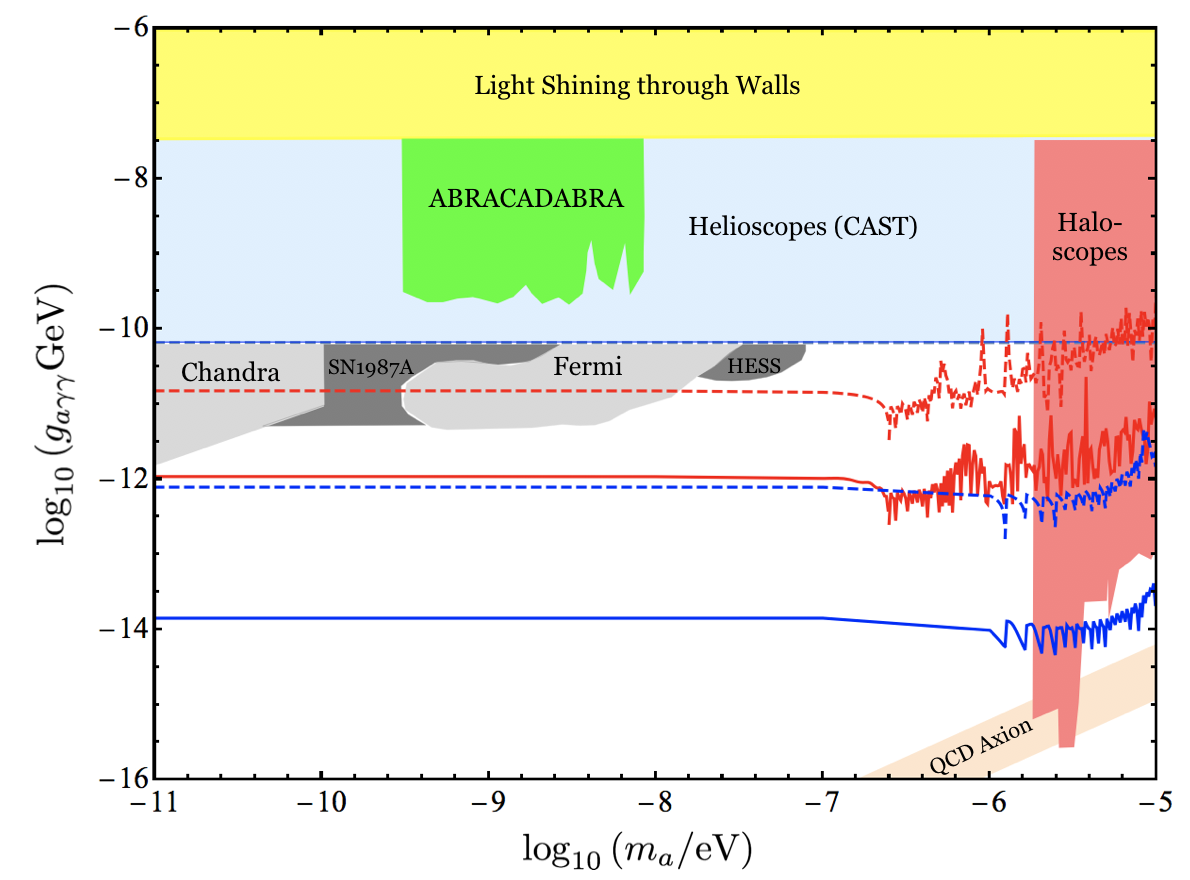}
    \caption{Projected reach of 100 hours of observation of magnetar SGR 1806--20 (Blue) and pulsar J1856--3754 (Red) with the Five-hundred-meter Aperture Spherical Telescope (FAST), assuming the line-of-sight is $0^\circ$ (solid) and $90^\circ$ (dashed) from the rotation axis (here assumed to coincide with the direction of the pulsar magnetic dipole moment). The search is broadband with bandwidth set by the fundamental gap oscillation frequency $\Delta \nu \approx 100$ MHz (SGR 1806--20), 6 MHz (J1856--3754). Relevant parameters describing the pulsars are listed in the table.} 
    \label{fig:sensitivity}
\end{figure}

The fundamental gap oscillation frequency of RX J1856--3754 is $\nu_1 = 6$ MHz, which is undetectable by terrestrial radio telescopes. Thus any detectable signal is generated by higher order gap oscillation harmonics. The SGRs listed in the table do not suffer from the same issue as $\nu_{1, 1900+14} \sim 60$ MHz, $\nu_{1,1806-20} \sim 100$ MHz.  For axion mass $m_a$, the highest signal flux comes from the $n$th harmonic where $n = \ceil{\text{max}(m_a/2\pi,\nu_\text{min})/\nu_1}$ and $\nu_\text{min}$ is the lowest frequency observed by the telescope or interferometer. We propose detecting the axion-induced signal over an observing bandwidth equal to the fundamental gap oscillation frequency of the pulsar being observed, that is $\delta \nu = \nu_1$. Thus, in contrast with axion dark matter searches that look for thin radio lines, we assume the signal generated at frequency $\nu_n$ ($n$ corresponding to the gap oscillation harmonic generating the signal), is spread out over the separation between lines. The sensitivity of this proposal, shown in Fig. \ref{fig:sensitivity}, corresponds to 100 hours of observation of RX J1856.5--3754 and SGR 1806--20 with the Five-hundred-meter Aperture Spherical Telescope (FAST) \cite{FAST}.

\section{Conclusions \label{sec:conclusions}}

In this work we presented a new mechanism for axion production in pulsar magnetospheres. As shown, oscillating gap regions, where $\bfE\cdot \bfB \ne 0$, can source axions (as in Light Shining through Walls experiments) with energies corresponding to those of the gap oscillation modes and whose abundance greatly exceeds that of axion dark matter near the neutron star. The conversion of the created axions to photons gives rise to enhanced radio emission at the Fourier frequencies of the gap oscillations. We found that 100 hours of observation of various Galactic pulsars with minimal radio emission could display an indirect signal of axion-photon conversion events. We found that radio missions operating at frequencies above 50 MHz, such as FAST and SKA, can probe axion-photon couplings that are orders of magnitude lower than astrophysical bounds over a wide range of masses. At high axion-photon couplings that are still not excluded by current bounds, axion-photon conversion events may be bright enough to explain the enigmatic fast radio bursts (FRBs) observed by a number of collaborations \cite{Lorimer2007, Thornton2013, Burke2014, Spitler2014, Ravi2015, Masui2015, Champion2016, Caleb2016, Bannister2017, Shannon2018, CHIME1, CHIME2}. The mechanism we have proposed is consistent with claims that FRBs have a magnetar origin \cite{Bochenek2020, Bhandari2020,Zhang2020}. {A more detailed analysis of the connection between gap oscillation-sourced axions and FRBs is the subject of ongoing investigation.} {The projected reach of this proposal also motivates the study of more sophisticated models of gap dynamics, such as those found in recent particle-in-cell simulations \cite{Spitkovsky2020}.}

\begin{acknowledgments}
\color{black} The author thanks Savas Dimopoulos for helpful conversations and Robert Lasenby for useful comments on signal detection and on the manuscript. The author also thanks Anatoly Spitkovsky and Roger Blandford for detailed discussion of vacuum gaps and radio emission properties of pulsars and magnetars. \color{black} This work was supported by the National Science Foundation under Grant No.
PHYS- 1720397 and the Gordon and Betty Moore Foundation Grant GBMF7946. The author
acknowledges the support of the Fletcher Jones Foundation and the National Science Foundation (NSF) Graduate Research Fellowship Program. 
\end{acknowledgments}

\clearpage

\vspace{5mm}

\bibliography{SparkGap}

\end{document}